\documentclass[graybox]{svmult}%
\usepackage{amssymb}
\usepackage{amsmath}
\usepackage{type1cm}
\usepackage{makeidx}
\usepackage{graphicx}
\usepackage{multicol}
\usepackage[bottom]{footmisc}
\usepackage{newtxtext}
\usepackage{newtxmath}
\usepackage{amsfonts}%
\providecommand{\U}[1]{\protect\rule{.1in}{.1in}}
\makeindex

\begin{document}

\title{Chiral Coupling to Magnetodipolar Radiation}

\institute{Tao Yu \at Max Planck Institute for the Structure and Dynamics of Matter, Luruper Chaussee 149, 22761 Hamburg, Germany, \email{tao.yu@mpsd.mpg.de}
	\and Gerrit E. W. Bauer \at Institute for Materials Research \& WPI-AIMR \& CSRN, Tohoku
	University, Sendai 980-8577, Japan,\\ Kavli Institute of
	NanoScience, Delft University of Technology, 2628 CJ Delft, The Netherlands,\\  \email{G.E.W.Bauer@tudelft.nl}}
\author{Tao Yu and Gerrit E. W. Bauer}
\maketitle




\abstract*{Each chapter should be preceded by an abstract (no more than 200
words) that summarizes the content. The abstract will appear \textit{online}
at \url{www.SpringerLink.com} and be available with unrestricted access. This
allows unregistered users to read the abstract as a teaser for the complete
chapter. Please use the 'starred' version of the \texttt{abstract} command for
typesetting the text of the online abstracts (cf. source file of this chapter
template \texttt{abstract}) and include them with the source files of your
manuscript. Use the plain \texttt{abstract} command if the abstract is also to
appear in the printed version of the book.}

\abstract{We review and extend the theory of chiral pumping of spin waves by
magnetodipolar stray fields that generate unidirectional spin currents
and asymmetric magnon densities. We illustrate the physical
principles by two kinds of chiral excitations of magnetic films, i.e., by the
evanescent Oersted field of a narrow metallic stripline with an AC current bias and
a magnetic nanowire under ferromagnetic resonance.}

\section{Introduction}

\label{sec:1} 

\textquotedblleft Handedness\textquotedblright\ or \textquotedblleft
chirality\textquotedblright\ of wave propagation is a lively research topic in
optics, acoustics, and condensed matter physics. The \textquotedblleft
spin\textquotedblright\ of magnons is rooted in the time-reversal symmetry
breaking of the magnetic order and leads to chiral coupling with other
excitations when locked to the momentum. This phenomenon is governed by
non-universal selection rules. This Chapter clarifies a specific mechanism,
viz. evanescent microwaves that can efficiently generate chiral dynamics. 

Magnonics and magnon spintronics
\cite{magnonics1,magnonics2,magnonics3,magnonics4} are emergent fields that
hold the promise of a next-generation low-power and scalable information
processing and communication technology. The generation of coherent and
propagating spin waves is a crucial ingredient, which can be realized by
magnetic fields generated by microwave antennas such as current-biased
metallic striplines. In order to generate stray fields with high-momentum
Fourier components these must be small in size and placed close to the
magnetic medium. Not only the amplitude, but also the direction of the excited
spin waves depend on the excitation conditions that obey right-hand rules and
are therefore \emph{chiral}.

In this Chapter, we focus on the chirality of the dipolar coupling between the
magnetization dynamics in ferromagnetic heterostructures
\cite{chiral_simulation,Yu1,Yu2,nanowire,accumulation1,accumulation2}, while
those in optics
\cite{chiral_review,chiral_optics1,chiral_optics2,chiral_optics3,chiral_optics4,chiral_optics5,chiral_emitter}%
, plasmonics \cite{near_field,Petersen}, and magnetic structures with
Dzyaloshinskii-Moriya interaction are treated in other chapters. We focus on
the favorite material of magnonics, viz. the ferrimagnetic insulator yttrium
iron garnet (YIG) with high Curie temperature and outstanding magnetic and
acoustic quality \cite{low_damping_nanometer}. Its magnons can be excited
electrically by heavy metal contacts \cite{Ludo}, acoustically \cite{acoustic}%
, as well as by a large spectrum of electromagnetic waves from gigahertz
(microwaves) to petahertz (light). Magnonic transducers with spatially
separated contact that excite and detect magnons
\cite{chiral_simulation,Ludo,centimeter,nanomagnetism,Dirk_2013,CoFeB_YIG,Co_YIG,Haiming_NC,Haiming_PRL}
are sensitive probes to study magnon transport. We illustrate the chiral
physics for thin YIG films with in-plane magnetizations, but other materials
and configurations can be treated by changing the model parameters.

The spin waves of in-plane magnetized films can be classified by the
interaction that governs their dispersion as a function of wave vector, into
the dipolar, dipolar-exchange and exchange type with energies ranging from a
few gigahertz to many terahertz
\cite{magnonics1,magnonics2,magnonics3,magnonics4,centimeter}. The
long-wavelength modes are dipolar, whereas the short-wavelength ones are
exchange. Bulk volume modes and surface (Damon-Eshbach) modes propagate along
or perpendicular to the magnetization direction with different dispersion
relations \cite{Walker,DE,spin_waves_book,new_book}. Moreover, the surface
modes are chiral: their propagation direction (linear momentum) is fixed by
the outer product of surface normal and magnetization direction, allowing
unidirectional spin current generation by dominantly exciting one surface of a
magnetic film
\cite{heat_conveyer1,heat_conveyer2,heat_conveyer3,heat_conveyer4}. However,
Damon-Eshbach spin waves are not well suited for applications --- their group
velocity tends to be zero when the linear momentum is larger than the inverse
of film thickness, leading to a small spin conductivity. They are also very
sensitive to dephasing by surface roughness \cite{surface_roughness}, and do
not exist in sufficiently thin films.

An alternative to intrinsically chiral spin waves is the chiral excitation of
non-chiral ones. Micromagnetic simulations \cite{chiral_simulation} revealed
that the AC dipolar field emitted by a magnetic nanowire on top of an in-plane
magnetized film with magnetization normal to the wire can excite
unidirectional spin waves. We have been motivated by experiments on an array
of magnetic nanowires on top of an ultrathin YIG film that generated
unidirectional spin waves parallel to the surface and perpendicular to the
nanowires \cite{Yu2} to develop a general theory of coherent and incoherent
chiral excitation of magnons \cite{Yu1,nanowire} by the dipolar interaction
between the dynamics of a magnetic film and a magnetic transducer. The
chirality can be traced to the different stray fields generated by spin waves
with opposite polarization and propagation. By angular momentum conservation
electromagnetic waves with particular polarization emitted by a magnetic
transducer couple only the circularly polarized component of a spin wave with
a certain propagation direction \cite{chiral_review}. When dipolar or crystal
anisotropy mixes the right and left circularly polarized components, magnons
are still excited preferentially, but not exclusively, in one direction.
Finally, a (short-range) exchange coupling between film and transducer is not
sensitive to the propagation direction, and reduces the chirality.\ 

Recently, two remarkable experiments confirmed our predictions. By NV
magnetometry Bertelli et al. \cite{NV_Teono} observed  chiral pumping of spin
waves by a stripline antenna. Wang et al.  \cite{two_wire_Haiming} measured
unidirectional microwave transmission mediated by two magnetic wires on top of
a thin magnetic film, i.e. chiral magnon-magnon coupling. 

The chiral coupling to spin waves enables the generation and control of spin
currents \cite{Yu1,Yu2,nanowire} or spin accumulations
\cite{accumulation1,accumulation2} in ferromagnetic insulators, which is
beneficial for spintronic devices. In this short review, we comprehensively
illustrate two kinds of chiral coupling to the magneto-dipolar radiation,
including the evanescent field of a thin stripline that carries an AC current
(Sec.~\ref{stripline}) and that of a magnetic wire under resonant excitation
(Sec.~\ref{dipolar}).

\section{Chiral excitation of spin waves by metallic stripline}

\label{stripline}
\begin{figure}[th]
\begin{center}
{\includegraphics[width=11.2cm]{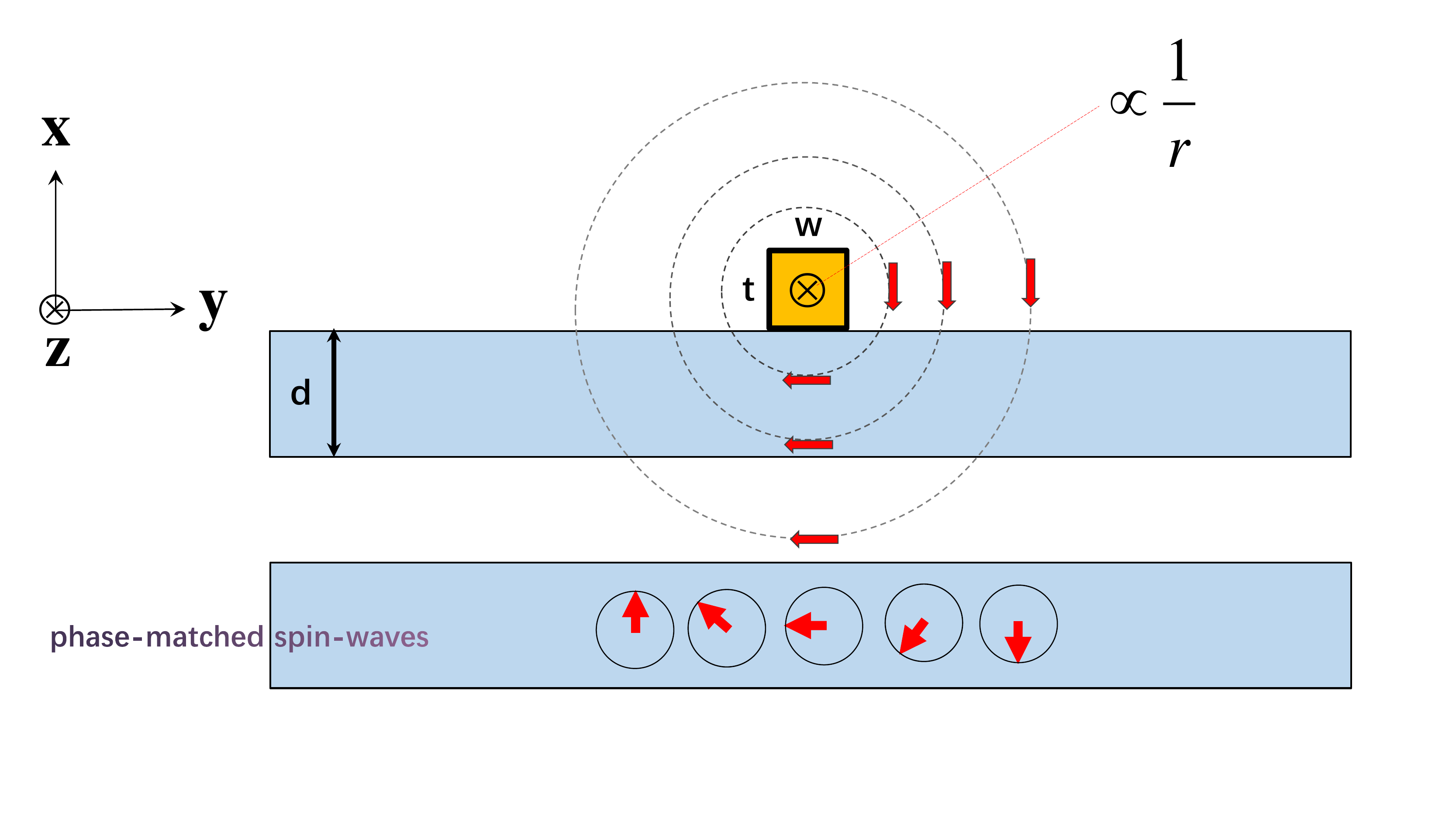}}
\end{center}
\caption{(Color online) Chiral excitation of spin waves in a magnetic thin
film by the near field of a stripline antenna. The ac magnetic field is
axially symmetric with an oscillating modulus and in the film a
position-dependent linear polarization. It excites spin waves with the same
frequency and phase-matched spatial amplitude. The film magnetization
direction (here parallel to the stripline) can be tuned by a static magnetic
field.}%
\label{simple_picture}%
\end{figure}

We call a wave \textquotedblleft chiral\textquotedblright\ when it propagates
with handedness, i.e. in a certain direction that is determined by two other
control vectors, such as surface normal and magnetic field. A rotating
\emph{electrical} dipole \cite{nano_optics,Jackson} excites surface plasmon
polaritons in one direction only \cite{near_field,Petersen,nano_optics}, while
a precessing \emph{magnetic} dipole excite magnons unidirectionally
\cite{poineering_1,poineering_2}. Here we analyze solutions of the combined
Maxwell and Landau-Lifshitz-Gilbert equations that explain the available
experimental evidence. We analyze the near microwave field from a normal metal
strip line in Sec.~\ref{Stripline} and its effect on a thin magnetic film in
Sec.~\ref{Magnetic_film} (see Fig.~\ref{simple_picture}). We focus for
simplicity on a configuration in which the film normal is along the
$x$-direction, $\hat{\mathbf{z}}$ is parallel to a stripline that is assumed
to be very long, and the excited spin waves propagate in the $y$-direction.

\subsection{Oersted magnetic fields\label{Stripline}}

We first demonstrate that even though the magnetic field of a stripline is
linearly-polarized in real space (see Fig. \ref{simple_picture}), it is chiral
in momentum space. Ampere's Law states that the current density $\mathbf{J}%
(\mathbf{r})$ generates the vector potential \cite{Jackson}
\begin{equation}
\mathbf{A}(\mathbf{r},t)=\frac{\mu_{0}}{4\pi}\int d\mathbf{r}^{\prime
}dt^{\prime}\frac{\mathbf{J}(\mathbf{r}^{\prime},t^{\prime})}{|\mathbf{r}%
-\mathbf{r}^{\prime}|}\delta\left(  t^{\prime}+\frac{|\mathbf{r}%
-\mathbf{r}^{\prime}|}{c}-t\right)  ,
\end{equation}
where $\mu_{0}$ is the vacuum permeability and the delta-function represents
(non-relativistic) retardation. For a harmonic source $\mathbf{J}\left(
t\right)  \sim\mathbf{J}(\omega)e^{-i\omega t}$,
\begin{equation}
\mathbf{A}(\mathbf{r},\omega)=\frac{\mu_{0}}{4\pi}\int d\mathbf{r}^{\prime
}\mathbf{J}(\mathbf{r}^{\prime},\omega)\frac{e^{ik|\mathbf{r}-\mathbf{r}%
^{\prime}|}}{|\mathbf{r}-\mathbf{r}^{\prime}|},
\end{equation}
where $k=\omega/c$. The current in the stripline is uniform over the cross
section of width $w$ and thickness $t$ as well as length $L\ll c/\omega.$ In
the long wavelength limit and square cross section $\mathbf{J}(\mathbf{r}%
,\omega)\simeq\delta(x)\delta(y)\mathcal{J}(\omega)\hat{\mathbf{z}},$ where
$\mathcal{J}$ is the total electric current, leading to%
\begin{equation}
\mathbf{A}(\mathbf{r},\omega)=\frac{\mu_{0}}{4\pi}\mathcal{J}(\omega
)\hat{\mathbf{z}}\int_{-\infty}^{\infty}dz^{\prime}\frac{e^{ik\sqrt
{x^{2}+y^{2}+z^{\prime2}}}}{\sqrt{x^{2}+y^{2}+z^{\prime2}}}, \label{A1}%
\end{equation}
which does not depend on $z$. Substituting the Weyl identity
\cite{nano_optics}
\begin{equation}
\frac{e^{ik\sqrt{x^{2}+y^{2}+z^{2}}}}{\sqrt{x^{2}+y^{2}+z^{2}}}=\frac{i}{2\pi
}\int dk_{y}dk_{z}\frac{e^{ik_{x}|x|+ik_{y}y+ik_{z}z}}{k_{x}}, \label{weyl}%
\end{equation}
where $k=\sqrt{k_{x}^{2}+k_{y}^{2}+k_{z}^{2}}$ and $k_{x}=|a|+i|b|$ is
complex, into Eq.~(\ref{A1}) yields
\begin{equation}
\mathbf{A}(x,y,\omega)=\frac{i\mu_{0}}{4\pi}\mathcal{J}(\omega)\hat
{\mathbf{z}}\int dk_{y}\frac{e^{ik_{x}|x|+ik_{y}y}}{k_{x}}.
\end{equation}
The magnetic field $\mathbf{H}(\mathbf{r})=\boldsymbol{\nabla}\times
\mathbf{A}(\mathbf{r})/\mu_{0}=\left(  \partial_{y}A_{z},-\partial_{x}%
A_{z},0\right)  /\mu_{0}$ is transverse to the wire, see Fig.
\ref{simple_picture}. Below the stripline $\left(  x<0\right)  $,
\begin{align}
H_{x}(x,y,\omega)  &  \equiv\int dk_{y}e^{ik_{y}y}H_{x}(x,k_{y})=-\int
dk_{y}e^{ik_{y}y}\frac{\mathcal{J}(\omega)}{4\pi}\frac{k_{y}}{k_{x}}%
e^{-ik_{x}x},\nonumber\\
H_{y}(x,y,\omega)  &  \equiv\int dk_{y}e^{ik_{y}y}H_{y}(x,k_{y})=-\int
dk_{y}e^{ik_{y}y}\frac{\mathcal{J}(\omega)}{4\pi}e^{-ik_{x}x},
\end{align}
where $k_{x}=\sqrt{(\omega/c)^{2}-k_{y}^{2}}.$ Directly above or below the
wire $H_{x}\left(  x,y=0,\omega\right)  =0,$ i.e. the magnetic field is
linearly-polarized along $y.$ The polarization rotates as a function of $y$
until $H_{y}\left(  0,y\rightarrow\infty,\omega\right)  =0.$ Surprisingly, a
circular polarization emerges in the Fourier components
\begin{align}
H_{x}(x,k_{y},\omega)  &  =-\frac{\mathcal{J}(\omega)}{4\pi}\frac{k_{y}}%
{k_{x}}e^{-ik_{x}x},\nonumber\\
H_{y}(x,k_{y},\omega)  &  =-\frac{\mathcal{J}(\omega)}{4\pi}e^{-ik_{x}x}.
\end{align}
For an evanescent field with $k_{y}>\omega/c\equiv k$, $k_{x}=i\sqrt{k_{y}%
^{2}-k^{2}}$ and $x<0$
\begin{align}
H_{x}(x,k_{y},\omega)  &  =\frac{i\mathcal{J}(\omega)}{4\pi}\frac{k_{y}}%
{\sqrt{k_{y}^{2}-k^{2}}}e^{\sqrt{k_{y}^{2}-k^{2}}x},\nonumber\\
H_{y}(x,k_{y},\omega)  &  =-\frac{\mathcal{J}(\omega)}{4\pi}e^{\sqrt{k_{y}%
^{2}-k^{2}}x}.
\end{align}

At microwave frequencies $\omega/(2\pi)\sim10$~GHz, $k\equiv\omega/c\sim
200$~m$^{-1}$ and wavelength $\lambda=2\pi/k\sim3$~cm. The spin wavelength at
the same frequency is much smaller with $\sqrt{k_{y}^{2}+k_{z}^{2}}\gg
\omega/c$, so we are in the near-field limit. The magnetic field component
$H_{x}\rightarrow i\mathrm{sgn}(k_{y})H_{y}$ is then circularly polarized with
a sign locked to its linear momentum.

For a finite rectangular cross section with $0<x<t$ and $-w/2<y<w/2$ the
Fourier components of the magnetic field read
\begin{align}
H_{x}(x,k_{y},\omega)  &  =i\frac{J(\omega)}{4\pi}\mathcal{F}(t,w)\frac{k_{y}%
}{\sqrt{k_{y}^{2}-k^{2}}}e^{\sqrt{k_{y}^{2}-k^{2}}x},\nonumber\\
H_{y}(x,k_{y},\omega)  &  =-\frac{J(\omega)}{4\pi}\mathcal{F}(t,w)e^{\sqrt
{k_{y}^{2}-k^{2}}x}, \label{stripline_field}%
\end{align}
which differ from the previous results only by the form factor
\begin{equation}
\mathcal{F}(t ,w)=\frac{4}{k_{x}k_{y}}e^{ik_{x}\frac{t}{2}}\sin\left(
k_{x}\frac{t}{2}\right)  \sin\left(  k_{y}\frac{w}{2}\right)  .
\end{equation}
Irrespective to the shape of the stripline, the magnetic field components are
circularly polarized when $\left\vert k_{y}\right\vert \gg\omega/c$ but
oscillate now as function of the wave vector.

\subsection{Chiral excitation of spin waves\label{Magnetic_film}}

We focus here on thin YIG films with thickness $d\sim\mathcal{O}%
(10~\mathrm{nm})$, which allows an analytical treatment of the dispersion and
spin wave amplitudes in the dipolar-exchange regime \cite{Yu1}. An applied
magnetic field $H_{\mathrm{app}}\hat{\mathbf{z}}$ parallel to the stripline
corresponds to the Damon-Eshbach configuration, but we stress that for
ultrathin films there are no Damon-Eshbach surface modes. The spin wave energy
dispersion \cite{Yu1}
\begin{equation}
\omega_{\mathbf{k}}=\mu_{0}\gamma M_{s}\sqrt{\left[  \Omega_{H}+\alpha
_{\mathrm{ex}}k^{2}+1-f(|k_{y}|)\right]  \left[  \Omega_{H}+\alpha
_{\mathrm{ex}}k^{2}+(k_{y}^{2}/k^{2})f(|k_{y}|)\right]  }, \label{dispersion}%
\end{equation}
where $-\gamma$ is the electron gyromagnetic ratio, $M_{s}$ denotes the
saturated magnetization, $\alpha_{\mathrm{ex}}$ is the exchange stiffness,
$\Omega_{H}\equiv H_{\mathrm{app}}/M_{s}$, and
\begin{equation}
f(|k_{y}|)=1-\frac{1}{|k_{y}|d}+\frac{1}{|k_{y}|d}\exp(-|k_{y}|d),
\end{equation}
is highly anisotropic. The spin waves amplitudes across sufficiently thin
films are constant \cite{Yu1}:
\begin{equation}
m_{x}=\sqrt{\frac{B+1}{4d(B-1)}},~~~~~m_{y}=i\sqrt{\frac{B-1}{4d(B+1)}},
\label{amplitudes}%
\end{equation}
where we chose the normalization \cite{Kittel_book,HP,surface_roughness}
\begin{equation}
\int d\mathbf{r}\left[  m_{x}(\mathbf{r})m_{y}^{\ast}(\mathbf{r})-m_{x}^{\ast
}(\mathbf{r})m_{y}(\mathbf{r})\right]  =-i/2,
\end{equation}
and
\begin{equation}
B=\frac{1/2-(1/2)\left(  1+k_{y}^{2}/k^{2}\right)  f(|k_{y}|)}{\omega
_{\mathbf{k}}/(\mu_{0}\gamma M_{s})-\left(  \Omega_{H}+\alpha_{\mathrm{ex}%
}k_{y}^{2}+1/2\right)  +(1/2)\left(  1-k_{y}^{2}/k^{2}\right)  f(|k_{y}|)}.
\end{equation}
When $k_{y}\rightarrow0$: $f(|{k_{y}}|)=0$, $\lim_{k_{y}\rightarrow0}%
\omega_{\mathbf{k}}=\mu_{0}\gamma M_{s}\sqrt{\Omega_{H}(\Omega_{H}+1)}$,
$B\rightarrow-1-2\Omega_{H}-2\sqrt{\Omega_{H}(\Omega_{H}+1)}$. When
$\Omega_{H}\rightarrow0$ with a small static magnetic field, $B\rightarrow
-1-2\sqrt{\Omega_{H}}$, $\left\vert m_{y}\right\vert \gg\left\vert
m_{x}\right\vert $, so the Kittel mode is (nearly) linearly polarized. In the
opposite (exchange) limit of$\ \left\vert k_{y}\right\vert d\gg1$ and
$\alpha_{\mathrm{ex}}k^{2}\gg1$, $f(k_{y})\rightarrow1$, $\left\vert
B\right\vert \gg1$, and the spin waves are right-circularly polarized with
$m_{y}=im_{x}$.

The Oersted magnetic fields from the stripline interact with spin waves by the
Zeeman interaction \cite{Landau}
\begin{equation}
\hat{H}_{\mathrm{int}}=-\mu_{0}\int\mathbf{M}(\mathbf{r})\cdot\mathbf{H}%
(\mathbf{r})dV.
\end{equation}
The excited magnetization in the film can be expressed by time-dependent
perturbation theory \cite{Mahan}
\begin{equation}
M_{\alpha}(x,\boldsymbol{\rho},t)=-i\int_{-\infty}^{t}dt^{\prime}\left\langle
\left[  \hat{\mathbf{M}}_{\alpha}(x,\pmb{\rho},t),\hat{H}_{\mathrm{int}%
}(t^{\prime})\right]  \right\rangle .
\end{equation}
in terms of the retarded spin susceptibility tensor
\begin{equation}
\chi_{\alpha\delta}(x,x^{\prime};\boldsymbol{\rho}-\boldsymbol{\rho}^{\prime
};t-t^{\prime})=i\Theta(t-t^{\prime})\left\langle \left[  \hat{\mathbf{S}%
}_{\alpha}(x,\boldsymbol{\rho},t),\hat{\mathbf{S}}_{\delta}(x^{\prime
},\boldsymbol{\rho}^{\prime},t^{\prime})\right]  \right\rangle ,
\end{equation}
where $\hat{\mathbf{S}}_{\alpha}=-\hat{\mathbf{M}}_{\alpha}/(\gamma\hbar) $ is
the spin operator and a sum over repeated indices is implied. Hence
\cite{Yu1,Yu2,nanowire},
\begin{equation}
\mathbf{M}_{\alpha}(x,k_{y},\omega)=\mu_{0}(\gamma\hbar)^{2}\int_{-d}%
^{0}dx^{\prime}\chi_{\alpha\beta}(x,x^{\prime},k_{y},\omega)H_{\beta
}(x^{\prime},k_{y},\omega),
\end{equation}
where
\begin{equation}
\chi_{\alpha\beta}(x,x^{\prime},\mathbf{k},\omega)=-\frac{2M_{s}}{\gamma\hbar
}m_{\alpha}^{\left(  \mathbf{k}\right)  }(x)m_{\beta}^{\left(  \mathbf{k}%
\right)  \ast}(x^{\prime})\frac{1}{\omega-\omega_{\mathbf{k}}+i\Gamma
_{\mathbf{k}}}.
\end{equation}
Here, $\Gamma_{\mathbf{k}}=2\alpha\omega_{\mathbf{k}}$ is the reciprocal
lifetime in terms of the Gilbert damping constant $\alpha$. The excitation
efficiency is determined by $m_{\beta}^{\left(  k_{y}\right)  \ast}(x^{\prime
})H_{\beta}(x^{\prime},k_{y},\omega)$, so the excitation of circularly
polarized spin waves is chiral (or unidirectional) by the
polarization-momentum locking with the stripline magnetic field. Since the
amplitudes across thin films are constant for $kd\ll1$, the excited
magnetization in time domain and position space is the real part of the
inverse Fourier transform ($q\equiv k_{y}$),
\begin{align}
\mathbf{M}_{\alpha}(x,y,t)  &  =\sum_{q}e^{iqy-i\omega t}\mathbf{M}_{\alpha
}(x,q)\nonumber\\
&  \approx2i\mu_{0}\gamma\hbar dM_{s}m_{\alpha}^{\left(  q_{\omega}\right)
}m_{\beta}^{\left(  q_{\omega}\right)  \ast}\frac{1}{v_{q_{\omega}}%
}e^{-i\omega t}\left\{
\begin{array}
[c]{c}%
e^{iq_{\omega}y-\delta_{\omega}y}H_{\beta}(q_{\omega},\omega)\\
e^{-iq_{\omega}y+\delta_{\omega}y}H_{\beta}(-q_{\omega},\omega)
\end{array}
\text{ for }%
\begin{array}
[c]{c}%
y>0\\
y<0
\end{array}
\right.  , \label{excitations}%
\end{align}
where $q_{\omega}+i\delta_{\omega}$ is the positive root of $\omega_{q}%
=\omega+i\Gamma_{q}$, and $v_{q_{\omega}}$ is the modulus of the group
velocity $|\partial\omega_{q}/\partial q|_{q_{\omega}}$. The
polarization-momentum locking of the stripline field generates two different
magnetization dynamics. When the excited spin waves are circularly polarized,
they not only propagate in one direction only, but the excitation is also
spatially limited to half of the film, i.e. the chirality is perfect. We can
understand this phenomenon in terms of the interference between the spin waves
and the stripline magnetic field that is constructive and destructive on
opposite sides, as illustrated in Fig.~\ref{simple_picture}.

The dominant excitation direction can be switched with the film magnetization.
For a finite angle $\theta$ between the saturated magnetization and the
stripline, the situation becomes complicated by the reduced symmetry. It is
advantageous to transform Eq.~(\ref{excitations}) following the Supplements of
Refs.~\cite{Yu2,nanowire}:
\[
m_{x}^{(k_{y})}\rightarrow m_{x}^{(\mathbf{l)}},~~~~m_{y}^{(k_{y})}%
\rightarrow\cos\theta m_{y}^{(\mathbf{l})}%
\]
where $\mathbf{l}=(0,q\cos\theta,q\sin\theta)$ and $q$ is determined by
$\omega_{\mathbf{l}}+i2\alpha\omega_{\mathbf{l}}=\omega$. Even for circularly
polarized spin waves, the chirality is not perfect anymore while situation is
complicated for elliptical spin waves since their polarization depends on the
wave vector. For $\theta_{c}=\pi/2$ the chirality always vanishes. Since
mirror symmetry is broken, the two roots $\left\vert q_{\omega}^{(+)}%
+i\delta_{\omega}^{(+)}\right\vert \neq\left\vert -q_{\omega}^{(-)}%
-i\delta_{\omega}^{(-)}\right\vert $ for $\theta\neq0,\theta_{c}$. The
wavelength and propagation direction of the excited spin waves may therefore
be different on the two sides of the stripline.

\begin{figure}[th]
\begin{center}
{\includegraphics[width=6.8cm]{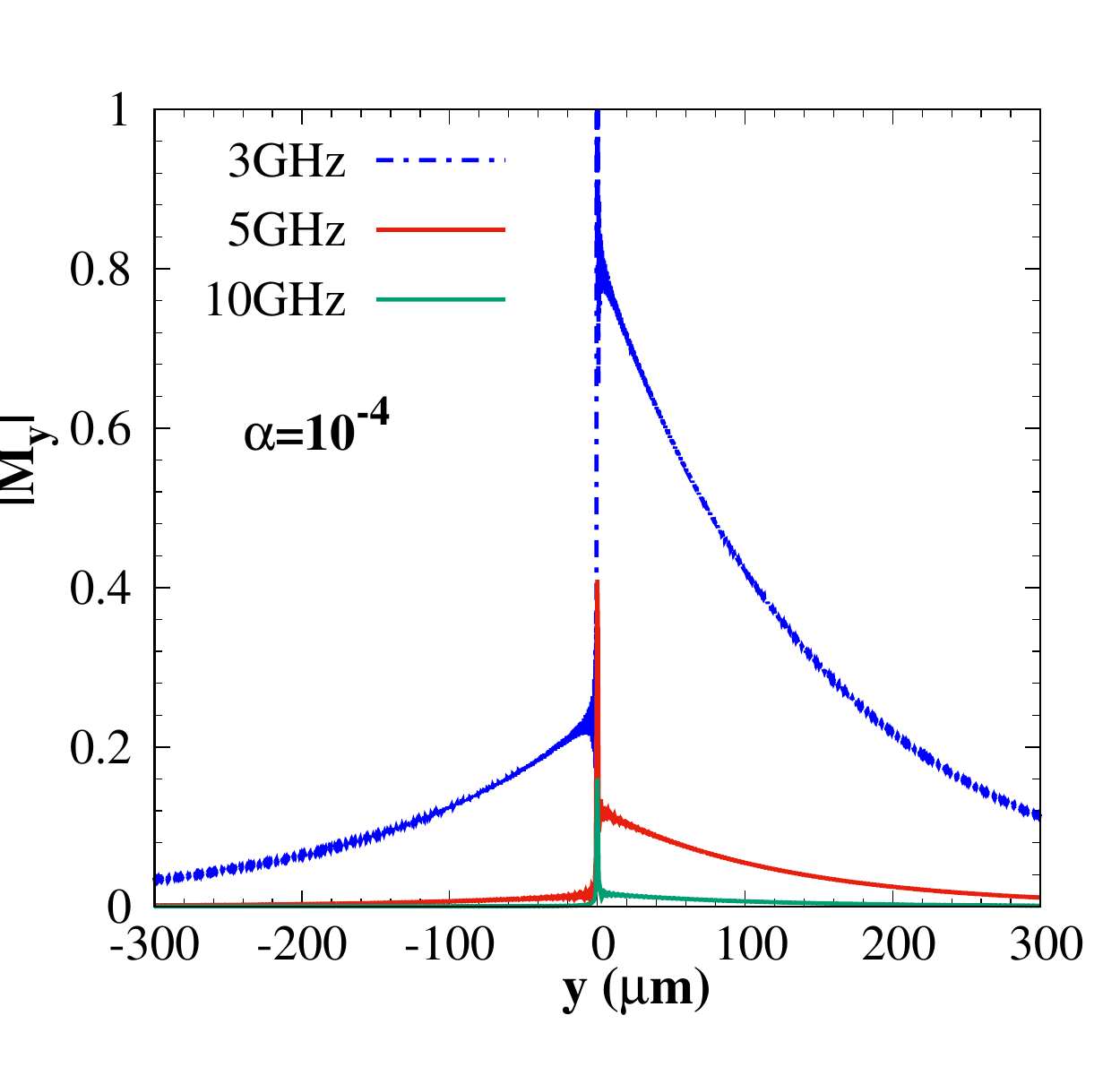}}
\end{center}
\caption{Calculated magnetization amplitude profile $\left\vert M_{y}\left(
y\right)  \right\vert $ of a YIG film with ground state magnetization along
$z,$ $d=20$~nm, and $\alpha=10^{-4},$ excited by a metal stripline with
$t=100$~nm, $w=1~\mathrm{\mu}$m, carrying an AC current with excitation
frequencies $\omega/(2\pi)=3$, 5, and 10~GHz. $\left\vert M_{y}\left(
y\right)  \right\vert $ is proportional to the current density, which is here
normalized by its maximum value for $\omega/(2\pi)=3$~GHz.}%
\label{magnetization_space}%
\end{figure}

Figure~\ref{magnetization_space} is a plot of the calculated excited
magnetization profile for a YIG\ magnetic film for constant current density
but different excitation frequencies $\omega/(2\pi)$. At low frequencies the excitation efficiency is high, but since
the dipolar interaction renders the spin wave precession elliptical, the
chirality is relatively weak. At high frequencies the chirality improves, but
the magnetization amplitude is suppressed by the form factor $\sin(k_{y}w/2)$
that favors spin waves with wavelengths around $w$. A narrower stripline helps
to excite spin waves with short wavelengths and higher chirality. The spatial
decay on both sides of the stripline is governed by the Gilbert damping.
Chiral spin waves can also be generated by magnetic striplines with high
coercivity that allow efficient excitation and almost perfect chirality at
frequencies $>10$~GHz. The physics is quite different, however, and explained
in the following section.

\section{Chiral spin wave excitation and absorption by a magnetic transducer}

\label{dipolar} 
\begin{figure}[th]
\begin{center}
{\includegraphics[width=11.2cm]{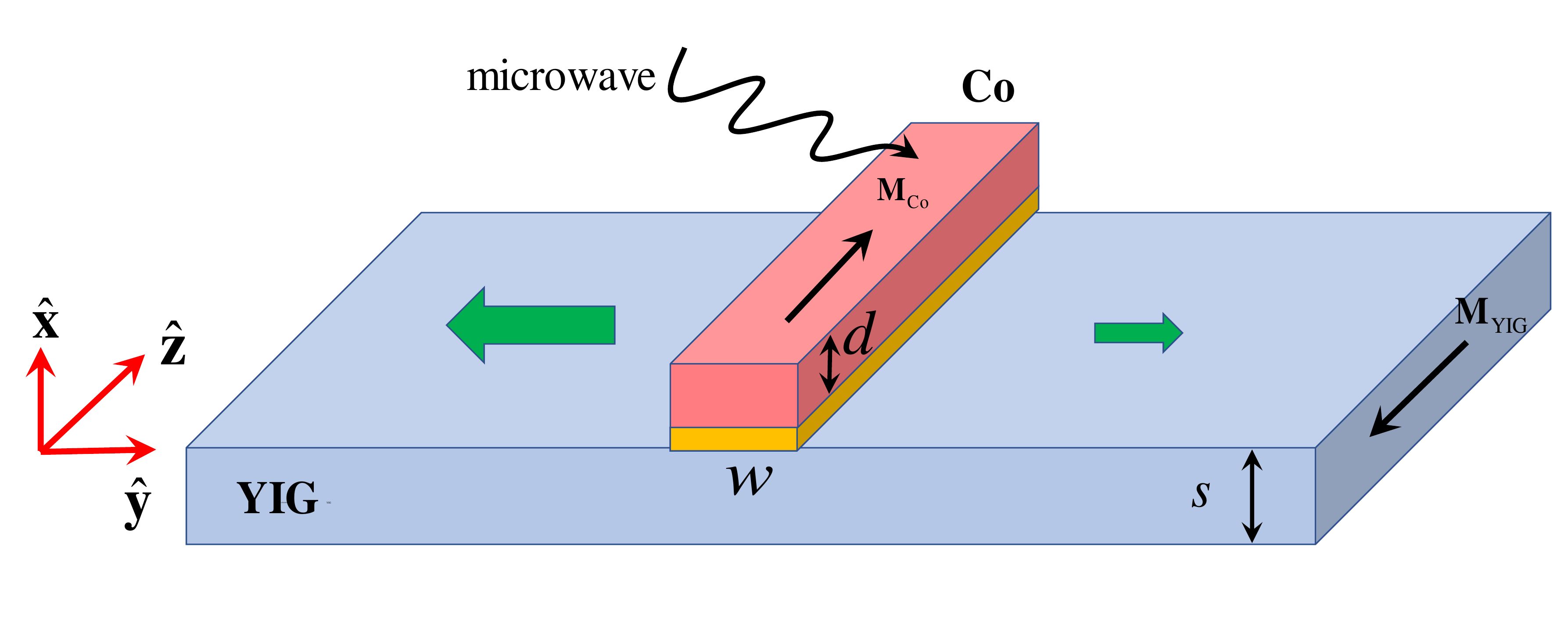}}
\end{center}
\caption{A magnetic (Co) nanowire transducer separated by a non-magnetic
spacer (optional) from a YIG film. The dipolar coupling is maximized for the
antiparallel magnetization. The direction of the magnon spin currents pumped
into the $\pm\hat{\mathbf{y}}$-directions is indicated by the green arrows,
whose size indicates the magnitude of the magnon currents. The black arrows
indicate the (nearly uniform) microwave input to the magnetic nanowire. }%
\label{model}%
\end{figure}

Coherent exchange-dipolar spin waves with short wavelengths
$\lambda<100$~nm are attractive information carriers by their long lifetime
and high group velocity. According to the discussion above their excitation is
difficult because striplines cannot be fabricated much finer than this wave
length. A small stripline cross section also increases Joule heating and
thereby limits the maximum applicable currents. A new strategy is to use
magnetic nanowires with high coercivity and resonance frequencies that can be
fabricated with the same feature sizes as normal metal ones. Rather than
applying an AC current directly, magnetic nanowires can be used as
\textquotedblleft antennas\textquotedblright\ that are excited by proximity
coplanar wave guides
\cite{chiral_simulation,Dirk_2013,CoFeB_YIG,Co_YIG,Haiming_NC,Haiming_PRL}. A
direct contact between film and nanowires can suppress chirality by the
interface exchange interaction and associated spin transfer \cite{Yu1}, but an
insulating spacer of a few atomic monolayers strongly suppresses exchange
without much affecting the dipolar interaction. Figure\ref{model} shows a
typical configuration with a Co nanowire on top of the YIG film.

\subsection{Chiral magnetodipolar field}

\begin{figure}[th]
\begin{center}
{\includegraphics[width=8.8cm]{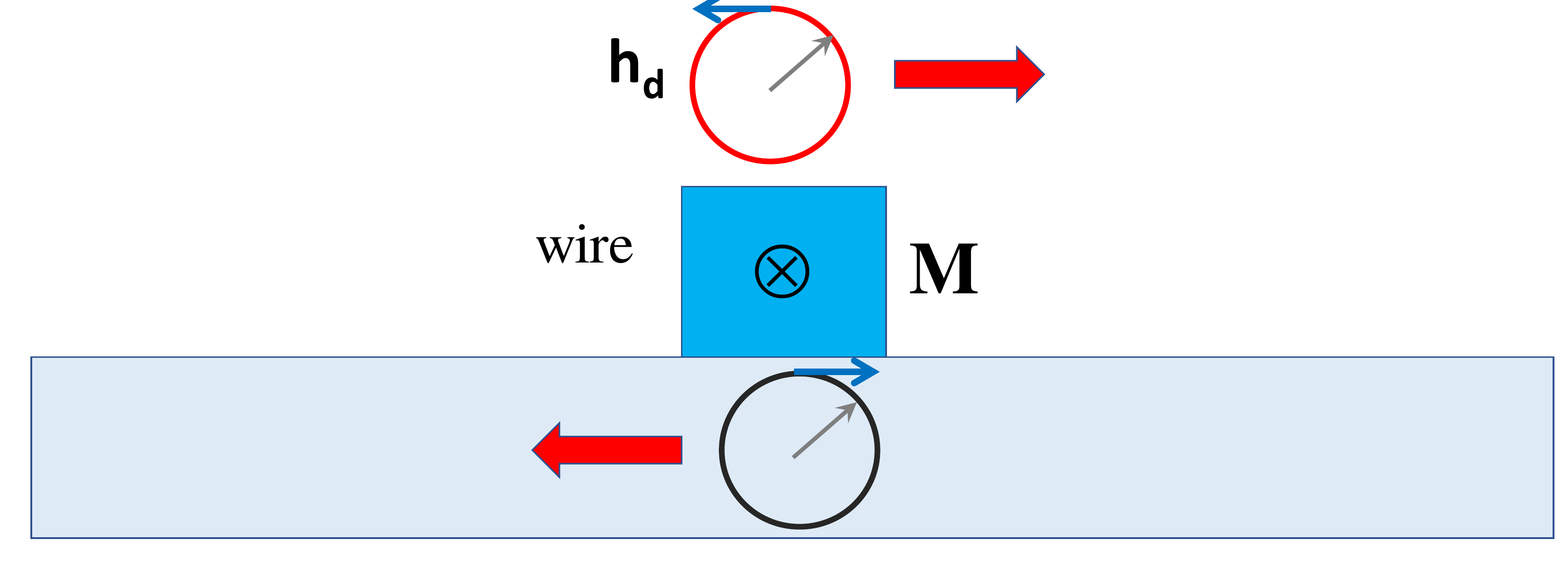}}
\end{center}
\caption{Dipolar magnetic field $\mathbf{\tilde{h}}$ generated by a Kittel
mode excitation of a magnetic nanowire ($\parallel\hat{\mathbf{z}}$). The
thick red and thin blue arrows indicate the propagation and precession
directions of $\mathbf{\tilde{h},}$ respecctively, both above and below the
wire.}%
\label{dipolar_field}%
\end{figure}

The dipolar field from the magnetic nanowire fundamentally differs from the
Oersted field of the AC current-biased normal metal wire discussed above. The
precessing magnetization is a magnetic dipole and generates a rotating dipolar
field rather than the oscillating axially symmetric field of the normal metal
wire sketched in Fig.~\ref{simple_picture}. The amplitudes of dipolar waves
decay faster than that of (monopolar) current-induced ones, but are still
long-ranged compared to e.g. the exchange interaction. The nanowire and its
equilibrium magnetization are parallel to the $z$-direction as shown in
Fig.~\ref{model}. When driven with a frequency $\omega$, the macrospin
(Kittel) magnetization dynamics of a wire with thickness $d$ and width $w$ is
the real part of
\begin{equation}
\tilde{M}_{x,y}(\mathbf{r},t)=\tilde{m}_{x,y}\Theta(x)\Theta(-x+d)\Theta
(y+w/2)\Theta(-y+w/2)e^{-i\omega t}, \label{Mxy}%
\end{equation}
where $\Theta(x)$ is the Heaviside step function and $\tilde{m}_{x,y}$ are
constant amplitudes that depend on the geometry and the excitation power. The
corresponding dipolar magnetic field \cite{Landau}
\begin{align}
\tilde{h}_{\beta}(\mathbf{r},t)  &  =\frac{1}{4\pi}\partial_{\beta}%
\partial_{\alpha}\int\frac{\tilde{M}_{\alpha}(\mathbf{r}^{\prime}%
,t)}{|\mathbf{r}-\mathbf{r}^{\prime}|}d\mathbf{r}^{\prime}\nonumber\\
&  =\frac{1}{4\pi}\partial_{\beta}\partial_{\alpha}\int dz^{\prime}\int%
_{0}^{d}dx^{\prime}\int_{-\frac{w}{2}}^{\frac{w}{2}}dy^{\prime}\frac{\tilde
{m}_{\alpha}e^{-i\omega t}}{\sqrt{z^{\prime2}+(x-x^{\prime})^{2}+(y-y^{\prime
})^{2}}}.
\end{align}
We use the Coulomb integral \cite{Yu1,nano_optics}
\begin{equation}
\frac{1}{\sqrt{z^{\prime2}+(x-x^{\prime})^{2}+(y-y^{\prime})^{2}}}=\frac
{1}{2\pi}\int dk_{x}dk_{y}\frac{e^{-|z^{\prime}|\sqrt{k_{x}^{2}+k_{y}^{2}}}%
}{\sqrt{k_{x}^{2}+k_{y}^{2}}}e^{ik_{x}(x-x^{\prime})+ik_{y}(y-y^{\prime})},
\end{equation}
a variation of the Weyl identity used in Eq.~(\ref{weyl}), to express the
magnetic field below the nanowire ($x<0$) with partial Fourier components
$k_{y}$
\begin{align}
\tilde{h}_{\beta}(k_{y},x,t)  &  =\int h_{\beta}(\mathbf{r},t)e^{-ik_{y}%
y}dy\nonumber\\
&  =\frac{1}{\pi}\int dk_{x}(k_{x}\tilde{m}_{x}+k_{y}\tilde{m}_{y})k_{\beta
}e^{ik_{x}x-i\omega t}\frac{1}{k_{x}^{2}+k_{y}^{2}}\frac{1-e^{-ik_{x}d}%
}{ik_{x}} \frac{\sin(k_{y}w/2)}{k_{y}}.
\end{align}
Closing the contour of the $k_{x}$ integral in the lower half of the complex
plane yields
\begin{equation}
\left(
\begin{array}
[c]{c}%
\tilde{h}_{x}(k_{y},x,t)\\
\tilde{h}_{y}(k_{y},x,t)
\end{array}
\right)  =-\frac{i}{4\pi}e^{\left\vert k_{y}\right\vert x}(1-e^{-\left\vert
k_{y}\right\vert d})\frac{2\sin(k_{y}w/2)}{k_{y}\left\vert k_{y}\right\vert
}\left(
\begin{array}
[c]{cc}%
\left\vert k_{y}\right\vert  & ik_{y}\\
ik_{y} & -\left\vert k_{y}\right\vert
\end{array}
\right)  \left(
\begin{array}
[c]{c}%
\tilde{m}_{x}\\
\tilde{m}_{y}%
\end{array}
\right)  e^{-i\omega t}. \label{tilde}%
\end{equation}
The perfectly right-circularly polarized wire dynamics of the Kittel mode in
rectangular wires ($\tilde{m}_{y}=i\tilde{m}_{x}$ when $w=d$) implies that the
Fourier components of $\mathbf{\tilde{h}}$ with $k_{y}>0$ vanish. The Fourier
component with $k_{y}<0$ is then perfectly left circularly polarized $\left(
\tilde{h}_{y}=-i\tilde{h}_{x}\right)  $. Above the nanowire, the magnetic
field direction and polarization are reversed, as sketched in
Fig.~\ref{dipolar_field}. The elliptical polarization of the Kittel mode in
rectangular nanowires breaks the perfect chirality. Analogous expressions can
be derived for arbitrarily shaped magnetic transducers such as discs, but
analytical expressions become complex or may not exist when the symmetry is reduced.

Equation~(\ref{excitations}) can be used also for magnetic fields
$\mathbf{\tilde{h}}$ generated by a magnetic transducer, i.e. Eq.
(\ref{tilde}), a left-circularly polarized dipolar field that propagates to
the left. An ellipticity of the spin waves in the film does not affect the
chirality since the excited magnetization still propagates to the left and
lives only in the left half-space, but it reduces the excitation efficiency.
The same holds when the Kittel mode in a rectangular nanowire is elliptical
and the spin waves in the film are circularly polarized. We illustrate these
conclusion below from different viewpoint.

Let us compare the dipolar stray fields $\mathbf{\tilde{h}}$ emitted by the
excited magnetic wire and $\mathbf{H}$ generated by a stripline as discussed
in the previous section. The main difference between these \textquotedblleft
Oersted" vs. \textquotedblleft dipolar" radiation is that the latter has
additional chirality that induces a circularly-polarized magnetic field in
real space, in contrast to the linearly-polarized magnetic field of the
former. Equation~(\ref{tilde}) can be summarized as $\tilde{h}_{x}%
\propto|k_{y}|(\tilde{m}_{x}+i\mathrm{sgn}(k_{y})\tilde{m}_{y})$ and
$\tilde{h}_{y}\propto ik_{y}(\tilde{m}_{x}+i\mathrm{sgn}(k_{y})\tilde{m}%
_{y}).$ $\tilde{h}_{y}=i\mathrm{sgn}(k_{y})\tilde{h}_{x}$ is the
polarization-momentum locking in reciprocal space, which is the same as that
of the evanescent Oersted field. However, the magnetic chirality affects
$\tilde{m}_{x}+i\mathrm{sgn}(k_{y})\tilde{m}_{y}$: for right
circularly-polarized (when $w=d$) $\tilde{m}_{y}=i\tilde{m}_{x}$,
$\mathbf{\tilde{h}}$ simply vanishes for positive $k_{y}$. Thus, the magnetic
field is unidirectional with linear momentum components normal to the wire
that are negative, which is more than just a locking between polarization and
momentum. $\mathbf{\tilde{h}}$ therefore couples chirally to spins with
arbitrary polarizations.

The Zeeman interaction $\sim\mathbf{M}\cdot\mathbf{\tilde{H}}$ between the
wire and film is governed as used above is completely equivalent to the
interaction $\sim\mathbf{\tilde{M}}\cdot\mathbf{h}$, where $\mathbf{\tilde{M}%
}$ is the wire magnetization and $\mathbf{h}$ the dipolar field generated by
the spin waves in the film. It is instructive to discuss the physics from this
second viewpoint. We assume again that the equilibrium wire magnetization is
fixed by the form anisotropy to the $z$-direction. A sufficiently soft\ film
magnetization can be rotated in the $x$-$z$ plane by an applied magnetic
field, but we address here only (anti)parallel magnetizations but general wave
propagation direction \cite{Yu2,nanowire}. We allow for the elliptical spin
wave polarization in the magnetostatic regime. At frequency $\omega$ and in
the coordinate system defined in Fig.~\ref{model} with in-plane wave vector
$\mathbf{k}=k_{y}\hat{\mathbf{y}}+k_{z}\hat{\mathbf{z}},$ we define
$M_{x}(\mathbf{r},t)=m_{R}^{\left(  \mathbf{k}\right)  }(x)\cos(\mathbf{k}%
\cdot\boldsymbol{\rho}-\omega t)$ and $M_{y}(\mathbf{r},t)\equiv
-m_{R}^{\left(  \mathbf{k}\right)  }(x)\sin(\mathbf{k}\cdot\boldsymbol{\rho
}-\omega t),$ where $m_{R}^{\left(  \mathbf{k}\right)  }(x)$ is the
time-independent amplitude into the film and $\boldsymbol{\rho}=y\hat
{\mathbf{y}}+z\hat{\mathbf{z}}.$ The dipolar field outside the film with
$\alpha,\beta=\{x,y,z\}$ \cite{Landau},
\begin{equation}
h_{\beta}(\mathbf{r},t)=\frac{1}{4\pi}\partial_{\beta}\partial_{\alpha}\int
d\mathbf{r}^{\prime}\frac{M_{\alpha}(\mathbf{r}^{\prime},t)}{|\mathbf{r}%
-\mathbf{r}^{\prime}|}, \label{dipolar_field_definition}%
\end{equation}
then reads
\begin{equation}
\left(
\begin{array}
[c]{c}%
h_{x}(\mathbf{r},t)\\
h_{y}(\mathbf{r},t)\\
h_{z}(\mathbf{r},t)
\end{array}
\right)  =\left(
\begin{array}
[c]{c}%
\left(  k+\eta k_{y}\right)  \cos\left(  \mathbf{k}\cdot\boldsymbol{\rho
}-\omega t\right) \\
\left(  \frac{k_{y}^{2}}{k}+\eta k_{y}\right)  \sin\left(  \mathbf{k}%
\cdot\boldsymbol{\rho}-\omega t\right) \\
k_{z}\left(  \frac{k_{y}}{k}+\eta\right)  \sin\left(  \mathbf{k}%
\cdot\boldsymbol{\rho}-\omega t\right)
\end{array}
\right)  \frac{1}{2}e^{-\eta kx}\int dx^{\prime}m_{R}^{\mathbf{k}}\left(
x^{\prime}\right)  e^{\eta kx^{\prime}}, \label{dipolar_film}%
\end{equation}
where $x>0$ ($x<-s$) indicates the dipolar field above (below) the film,
$\eta=1$ ($-1$) when $x>0$ ($x<-s$), $k=\left\vert \mathbf{k}\right\vert $,
and the spatial integral is over the film thickness.

When $k_{z}=0$, $k_{y}\neq0$ spin waves propagate normal to the wire and
$h_{z}=0$. The distribution of the dipolar field above and below the film then
strongly depends on the sign of $k_{y}$: the dipolar field generated by the
right (left) moving spin waves only appears above (beneath) the film
\cite{Yu1,Yu2,nanowire} and precesses in the opposite direction of the
magnetization. These features provide an alternative explanation of the chiral
coupling between these spin waves and \emph{any} magnet close to the film
surface \cite{Yu1,Yu2}. The chiral dipolar coupling is most pronounced when
the magnetizations of the film and wire are antiparallel
\cite{Yu1,Yu2,nanowire}.

When the film magnetization is rotated by 90 degrees in perpendicular to the
wire, the wire magnetization excites spin waves that propagate
\textit{parallel} to the magnetization ($k_{y}=0$, $h_{y}=0$), which for thick
films correspond to the backward moving bulk modes. Surprisingly, these also
couple chirally to the wire dynamics, but by a different mechanism. According
to Eq.~(\ref{dipolar_field}), $h_{x}\propto\left\vert k_{z}\right\vert
\cos\left(  k_{z}z-\omega t\right)  $ and $h_{z}\propto\eta k_{z}\sin\left(
k_{z}z-\omega t\right)  $. The dipolar fields generated by spin waves with
positive (negative) $k_{z}$ are left (right) circularly polarized,
respectively, while below the film, the polarizations are reversed. These spin
waves chirally interact with the transducer magnet since the polarization of
the transverse magnetization dynamics of the latter has to match that of the
stray field $\mathbf{h}$ \cite{chiral_simulation}.

Therefore, two mechanisms contribute to the chiral excitation, depending on
the magnetic configuration. When spin waves propagate perpendicular to the
magnetization with opposite momenta, their dipolar fields vanish on opposite
sides of the film; when propagating parallel to the magnetization, their
dipolar field is chiral, i.e., polarization-momentum locked. Purely chiral
coupling between magnons can be achieved in the former case without
constraints on the polarization of the local magnet, but in the latter case
elliptical polarization of the wire leads to partial chirality.

The resonance frequency of a magnetic nanowire can be tuned by an applied
magnetic field and excites spin waves in a frequency window that is governed
by the wire form factor. The magnetodipolar field emitted by a coherently
excited magnetic nanowire array can also be chiral \cite{Yu1,Yu2}. However,
such a nanowire grating with period $a$ and translational symmetry
$na\hat{\mathbf{y}}$ excites discrete spin waves with momenta $(m\pi
/a)\hat{\mathbf{y}}$, where $\{m,n\}\in Z_{0}$ that are observable as sharp
and intense feature in the microwave transmission (more details are shown
below).\textit{\ }

\subsection{Non-local detection}

Here we illustrate the principle of non-local excitation and detection of
magnons by a device consisting of two magnetic nanowires on top of a YIG film.
The generation of DC currents by AC forces in the absence of a DC bias is
generally referred to as \textquotedblleft pumping\textquotedblright%
\ \cite{pumping_Buttiker}. Spin pumping is the injection of a spin current by
the magnetization dynamics of a magnet into a normal metal contact by the
interface exchange interaction \cite{spin_pumping_electron,non_local}.
C\emph{hiral spin pumping} is the generation of unidirectional spin waves by
the dynamics of a proximity magnetic wire as discussed above. Its inverse is
the \emph{chiral spin absorption, }i.e. the wire dynamics induced by the stray
fields caused by spin waves in the film. We develop below a semi-analytic
theory of chiral spin pumping/absorption for antiparallel magnetic
configurations and describe two effects --- non-reciprocal microwave
transmission and chiral spin Seebeck effect. Whereas the former is due to
coherent pumping by applied microwaves, the latter represents the incoherent
(thermal) pumping by a temperature difference
\cite{Spin_seebeck_exp,Spin_seebeck_theory1,Spin_seebeck_theory2,Spin_caloritronics}%
. Both effects can be observed in terms of the magnon population or
temperature in the detector, e.g., inductively or by light scattering.

We switch from a purely classical picture of previous sections to a quantum
description of the chiral coupling in terms of Hamiltonian matrix elements
between generalized harmonic oscillators. This does not introduce new physics
since we can simply replace operators by classical amplitudes, but it provides
a compact formalism used in many other fields such as nanomechanical systems
and optics, and prepares the stage for the treatment of real quantum problems.
For simplicity, we focus on the antiparallel magnetic configuration with
maximized dipolar coupling (for arbitrary magnetization directions see
\cite{nanowire}). The dipolar coupling of the wire magnetization
$\mathbf{\tilde{M}}$ with that of a film $\mathbf{M}$ is governed by the
Zeeman interaction with the respective stray magnetic fields $\mathbf{h}$ and
$\mathbf{\tilde{h}}$ \cite{Landau}
\begin{equation}
\hat{H}_{\mathrm{int}}/\mu_{0}=-\int\mathbf{\tilde{M}}(\mathbf{r}%
,t)\cdot\mathbf{h}(\mathbf{r},t)d\mathbf{r}=-\int\mathbf{M}(\mathbf{r}%
,t)\cdot\mathbf{\tilde{h}}(\mathbf{r},t)d\mathbf{r}, \label{Hint}%
\end{equation}
where $\mathbf{h}$ and $\mathbf{\tilde{h}}$ have been introduced in
Eqs.~(\ref{dipolar_film}) and (\ref{tilde}). The magnetization dynamics of
film ($\mathbf{\hat{M}}$) and nanowire ($\mathbf{\hat{\tilde{M}}}$) are now
interpreted as operators with Cartesian components $\beta\in\left\{
x,y\right\}  $. To leading order of the expansion in magnon creation and
annihilation operators \cite{Kittel_book,HP,surface_roughness},
\begin{align}
\hat{M}_{\beta}(\mathbf{r})  &  =-\sqrt{2M_{s}\gamma\hbar}\sum_{\mathbf{k}%
}\left[  m_{\beta}^{\left(  \mathbf{k}\right)  }(x)e^{i{\mathbf{k}}%
\cdot\pmb{\rho}}\hat{\alpha}_{\mathbf{k}}+\mathrm{H.c.}\right]  ,\nonumber\\
\hat{\tilde{M}}_{\beta}(\mathbf{r})  &  =-\sqrt{2\tilde{M}_{s}\gamma\hbar}%
\sum_{k_{z}}\left[  \tilde{m}_{\beta}^{(k_{z})}(x,y)e^{ik_{z}z}\hat{\beta
}_{k_{z}}+\mathrm{H.c.}\right]  , \label{expansion}%
\end{align}
where $M_{s}$ and $\tilde{M}_{s}$ are the respective saturation
magnetizations, $m_{\beta}^{\left(  \mathbf{k}\right)  }(x)$ and $\tilde
{m}_{\beta}^{(k_{z})}(x,y)$ are the spin wave amplitudes across the film and
nanowire, and $\hat{\alpha}_{\mathbf{k}}$ and $\hat{\beta}_{k_{z}}$ denote the
magnon (annihilation) operator in the film and nanowire, respectively. The
total system Hamiltonian then reads
\begin{align}
\hat{H}/\hbar &  =\sum_{\mathbf{k}}\omega_{\mathbf{k}}\hat{\alpha}%
_{\mathbf{k}}^{\dagger}\hat{\alpha}_{\mathbf{k}}+\sum_{k_{z}}\tilde{\omega
}_{k_{z}}\hat{\beta}_{k_{z}}^{\dagger}\hat{\beta}_{k_{z}}\nonumber\\
&  +\sum_{\mathbf{k}}\left(  g_{\mathbf{k}}e^{-ik_{y}y_{0}}\hat{\alpha
}_{\mathbf{k}}^{\dagger}\hat{\beta}_{k_{z}}+g_{\mathbf{k}}^{\ast}%
e^{ik_{y}y_{0}}\hat{\beta}_{k_{z}}^{\dagger}\hat{\alpha}_{\mathbf{k}}\right)
,
\end{align}
where $\omega_{\mathbf{k}}$ and $\tilde{\omega}_{k_{z}}$ are the frequencies
of spin waves in the film and nanowire and the coupling
\begin{equation}
g_{\mathbf{k}}=F(\mathbf{k})\left(  m_{x}^{(\mathbf{k})\ast},m_{y}%
^{(\mathbf{k})\ast}\right)  \left(
\begin{array}
[c]{cc}%
|\mathbf{k}| & ik_{y}\\
ik_{y} & -k_{y}^{2}/|\mathbf{k}|
\end{array}
\right)  \left(
\begin{array}
[c]{c}%
\tilde{m}_{x}^{(k_{z})}\\
\tilde{m}_{y}^{(k_{z})}%
\end{array}
\right)  , \label{coupling_constant}%
\end{equation}
with $F(\mathbf{k})=-\mu_{0}\gamma\sqrt{M_{s}\tilde{M}_{s}/L}\phi\left(
\mathbf{k}\right)  $. The form factor $\phi\left(  \mathbf{k}\right)
=2\sin(k_{y}w/2)(1-e^{-kd})(1-e^{-ks})/(k_{y}{k^{2}})$ couples spin waves with
wavelengths of the order of the nanowire width (mode selection) and
$\lim_{\mathbf{\ k}\rightarrow0}\phi\left(  \mathbf{k}\right)  =wsd$. Pure
exchange waves are right-circularly polarized with $m_{y}^{(k_{y})}%
=im_{x}^{(k_{y})}$ and their coupling is perfectly chiral since $g_{-|k_{y}%
|}=0$ and $g_{|k_{y}|}\neq0$.

\begin{figure}[th]
\begin{center}
{\includegraphics[width=9cm]{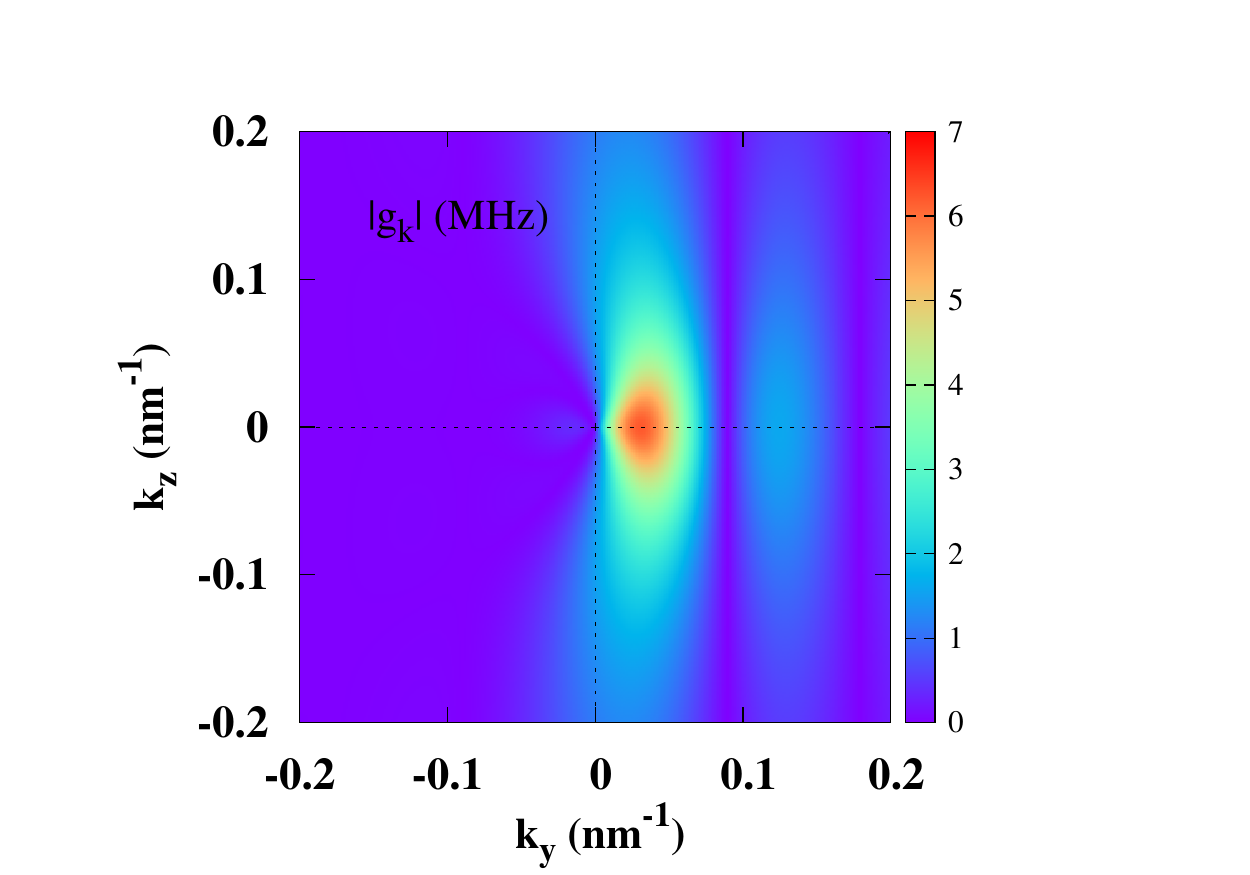}}
\end{center}
\caption{Momentum dependence of the dipolar coupling strength $|g_{\mathbf{k}%
}|$ of a magnetic nanowire and film (parameters in the text)~\cite{nanowire}.}%
\label{coupling}%
\end{figure}

Eqs.~(\ref{dispersion}) and (\ref{amplitudes}) give the spin-wave dispersion
and amplitudes in the thin film. The spin waves propagate in the nanowire
along $\hat{\mathbf{z}}$ with amplitudes \cite{Yu1,nanowire}
\begin{equation}
\tilde{m}_{x}^{k_{z}}=\sqrt{\frac{1}{4\mathcal{D}(k_{z})wd}},~~~~\tilde{m}%
_{y}^{k_{z}}=i\sqrt{\frac{\mathcal{D}(k_{z})}{4wd}}, \label{nanowire_waves}%
\end{equation}
where
\begin{equation}
\mathcal{D}(k_{z})=\sqrt{\frac{H_{\mathrm{app}}+N_{xx}\tilde{M}_{s}%
+\tilde{\lambda}_{\mathrm{ex}}k_{z}^{2}\tilde{M}_{s}}{H_{\mathrm{app}}%
+N_{yy}\tilde{M}_{s}+\tilde{\lambda}_{\mathrm{ex}}k_{z}^{2}\tilde{M}_{s}}}.
\end{equation}
$H_{\mathrm{app}}$ and $\tilde{\lambda}_{\mathrm{ex}}$ are the applied
magnetic field and the exchange stiffness of the nanowire, respectively. The
demagnetization factors $N_{xx}\simeq w/(d+w)$ and $N_{yy}=d/(d+w)$ \cite{Yu1}
also govern the spin waves frequency
\begin{equation}
\tilde{\omega}_{k_{z}}=\mu_{0}\gamma\sqrt{(H_{\mathrm{app}}+N_{yy}\tilde
{M}_{s}+\tilde{\lambda}_{\mathrm{ex}}k_{z}^{2}\tilde{M}_{s})(H_{\mathrm{app}%
}+N_{xx}\tilde{M}_{s}+\tilde{\lambda}_{\mathrm{ex}}k_{z}^{2}\tilde{M}_{s})}.
\label{omegaK}%
\end{equation}
When the magnetic field is antiparallel to the nanowire magnetization we
require $\left\vert H_{\mathrm{app}}\right\vert <\min\{N_{yy}\tilde{M}%
_{s},N_{xx}\tilde{M}_{s}\}$. The ellipticity of the Kittel mode with $k_{z}=0
$ is strongly affected by the shape anisotropy when the applied field is
sufficiently small and the aspect ratio large: when $d\ll w$, $N_{xx}%
\rightarrow1$, $N_{yy}\rightarrow0$, $\mathcal{D}$ is large and the mode is
nearly linearly-polarized. On the other hand, when $d\approx w$,
$\mathcal{D}\rightarrow1$, and the Kittel mode is circularly polarized. When
$d\lesssim w$, and the Kittel mode traces an elliptical orbit.
Figure~\ref{coupling} illustrates\ the chirality of the coupling parameter
$g_{\mathbf{k}}$ of the $k_{z}$-Kittel mode in a nanowire of dimensions
$w=70$~nm and $d=20$~nm and magnons in a film of thickness $s=20$~nm with wave
vector $\mathbf{k}=\left(  0,k_{y},k_{z}\right)  $ \cite{nanowire}. The
coupling maximum can be shifted to larger momenta by a smaller feature size of
the wire. The excitation of such short-wavelength spin waves is possible with
a magnetically hard transducer that has a high ferromagnetic resonance
frequency
\cite{chiral_simulation,Dirk_2013,CoFeB_YIG,Co_YIG,Haiming_NC,Haiming_PRL}.

\subsection{Coherent chiral spin wave transmission}

The quantum description leads to expressions that are fully equivalent with
Eq.~(\ref{excitations}) obtained from the classical description
\cite{input_output1,input_output2}. The excitation of magnons saps nanowire
energy and angular momentum, thereby contributing to the magnetization
damping, which can be observed as an increased linewidth of the ferromagnetic
resonance spectrum. In the quantum description, this broadening is determined
by the imaginary part of the magnetic self-energy, which in the first Born
approximation or the Fermi-golden rule reads
\begin{equation}
\delta\tilde{\kappa}_{k_{z}}=2\pi\sum_{k_{y}}|g_{\mathbf{k}}|^{2}\delta
(\tilde{\omega}_{k_{z}}-\omega_{\mathbf{k}}).
\end{equation}
We predict a very significant additional damping for a Co nanowire with width
$w=70$~nm, thickness $d=20$~nm, magnetization $\mu_{0}\tilde{M}_{s}=1.1$~T
\cite{Yu2,Haiming_PRL}, and exchange stiffness $\tilde{\lambda}_{\mathrm{ex}%
}=3.1\times10^{-13}$~cm$^{2}$ \cite{Co_exchange}. We adopt a YIG film
$s=20$~nm with magnetization $\mu_{0}M_{s}=0.177$~T and exchange stiffness
$\lambda_{\mathrm{ex}}=3.0\times10^{-12}$~cm$^{2}$
\cite{Yu2,Haiming_PRL,surface_roughness}. A magnetic field $\mu_{0}%
H_{\mathrm{app}}=0.05$~T is sufficient to switch the film magnetizations
antiparallel to that of the wire to maximize the effect
\cite{Haiming_NC,Haiming_PRL}. The calculated additional damping of nanowire
Kittel dynamics is then $\delta\alpha_{\mathrm{Co}}=\delta\tilde{\kappa
}_{k_{z}=0}/(2\tilde{\omega}_{k_{z}=0})=3.1\times10^{-2}$, which is one order
of magnitude larger than the intrinsic Gilbert damping coefficient
$\alpha_{\mathrm{Co}}=2.4\times10^{-3}$ \cite{Co_damping}.

\begin{figure}[th]
\begin{center}
{\includegraphics[width=8.0cm]{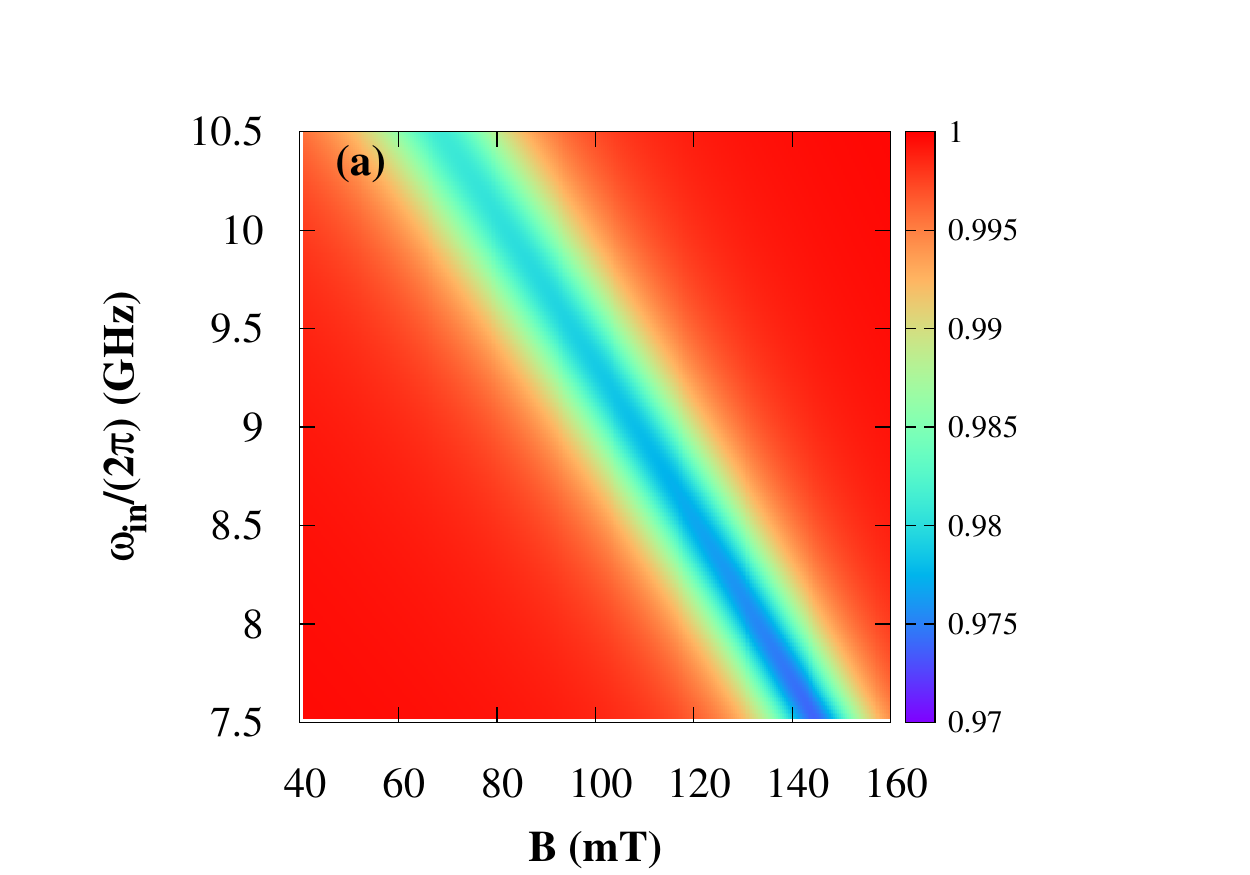}} \hspace{-0.4cm}
{\includegraphics[width=8.0cm]{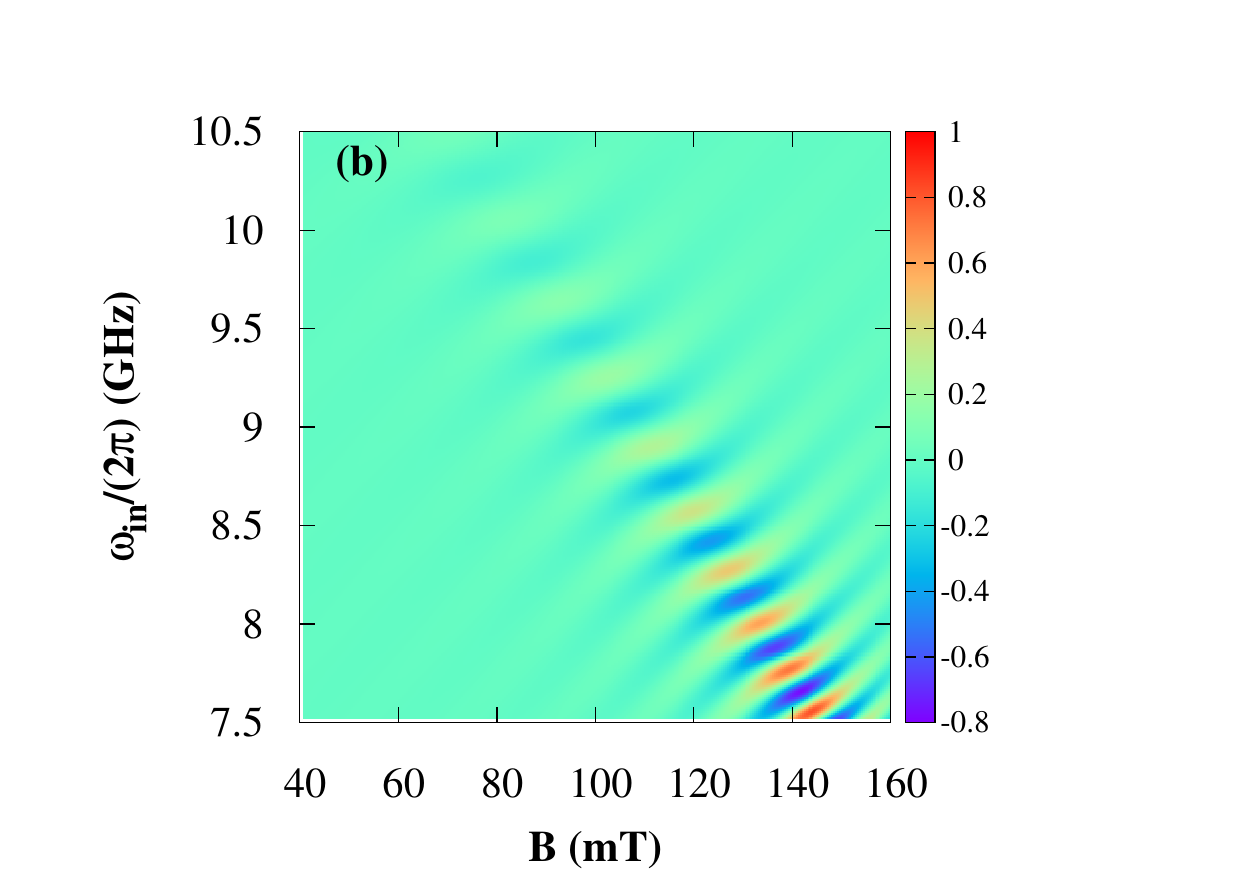}}
\end{center}
\caption{Microwave reflection $\mathrm{Re}S_{11}$ [(a)] and transmission
$\mathrm{Re}S_{12}$ [(b)] amplitudes, Eqs. (\ref{S}), for a system of two Co
nanowires on a YIG film as a function of frequency $\omega_{\mathrm{in}}$. The
radiative damping of both nanowires is $\kappa_{p}/(2\pi)=10$~MHz and other
parameters are given in the text.}%
\label{transmission}%
\end{figure}

Almost perfect chiral pumping by a nanowire \emph{array} has been observed by
the microwave transmission and Brillouin light scattering in Ref.~\cite{Yu2}.
We here focus on the new features in the broadband non-local
excitation-detection by \emph{two nanowires}. The magnetic order in two
nanowires located at $\mathbf{r}_{1}=R_{1}\hat{\mathbf{y}}$ and $\mathbf{r}%
_{2}=R_{2}\hat{\mathbf{y}}$ act as transducers for microwaves that are emitted
or absorbed by local microwave (normal metal) antennas such as coplanar wave
guides. The observable is the scattering matrix of the microwaves with
excitation (input) at $R_{1}$ and the detection (output) at $R_{2}$, which can
be formulated by the input-output theory \cite{input_output1,input_output2}.
The equation of motion of magnons localized at $R_{1}$ and $R_{2}$ with
operators $\hat{m}_{L}$ and $\hat{m}_{R}$ and coupled by the film magnons with
operators $\hat{\alpha}_{q}$ (not to be confused with the Gilbert damping
constant) read
\begin{align}
\frac{d\hat{m}_{L}}{dt}  &  =-i\omega_{\mathrm{K}}\hat{m}_{L}(t)-i\sum
_{q}g_{q}e^{iqR_{1}}\hat{\alpha}_{q}(t)-\left(  \frac{\kappa_{L}}{2}%
+\frac{\kappa_{p,L}}{2}\right)  \hat{m}_{L}(t)-\sqrt{\kappa_{p,L}}\hat
{p}_{\mathrm{in}}^{(L)}(t),\nonumber\\
\frac{d\hat{m}_{R}}{dt}  &  =-i\omega_{\mathrm{K}}\hat{m}_{R}(t)-i\sum
_{q}g_{q}e^{iqR_{2}}\hat{\alpha}_{q}(t)-\frac{\kappa_{R}}{2}\hat{m}%
_{R}(t),\nonumber\\
\frac{d\hat{\alpha}_{q}}{dt}  &  =-i\omega_{q}\hat{\alpha}_{q}(t)-ig_{q}%
e^{-iqR_{1}}\hat{m}_{L}(t)-ig_{q}e^{-iqR_{2}}\hat{m}_{R}(t)-\frac{\kappa_{q}%
}{2}\hat{\alpha}_{q}(t).
\end{align}
Here, $\kappa_{L}$ and $\kappa_{R}$ are the intrinsic damping of the Kittel
modes in the left and right nanowires, respectively, $\kappa_{p,L}$ is the
additional radiative damping induced by the microwave photons $\hat
{p}_{\mathrm{in}}^{(L)}$, i.e. the coupling of the left nanowire with the
microwave source, and $\kappa_{q}$ denotes the intrinsic (Gilbert) damping of
magnons in the films. In frequency space:
\begin{align}
\hat{\alpha}_{q}(\omega)  &  =g_{q}G_{q}\left(  \omega\right)  \left[
e^{-iqR_{1}}\hat{m}_{L}(\omega)+e^{-iqR_{2}}\hat{m}_{R}(\omega)\right]
,\nonumber\\
\hat{m}_{R}(\omega)  &  =\frac{-i\sum_{q}g_{q}^{2}G_{q}\left(  \omega\right)
e^{iq(R_{2}-R_{1})}}{-i(\omega-\omega_{\mathrm{K}})+\kappa_{R}/2+i\sum
_{q}g_{q}^{2}G_{q}\left(  \omega\right)  }\hat{m}_{L}(\omega),\nonumber\\
\hat{m}_{L}(\omega)  &  =\frac{-\sqrt{\kappa_{p,L}}}{-i(\omega-\omega
_{\mathrm{K}})+(\kappa_{L}+\kappa_{p,L})/2+i\sum_{q}g_{q}^{2}G_{q}\left(
\omega\right)  -f(\omega)}\hat{p}_{\mathrm{in}}^{(L)}(\omega),
\end{align}
with spin wave propagator $G_{q}\left(  \omega\right)  =\left[  (\omega
-\omega_{q})+i\kappa_{q}/2\right]  ^{-1}$ and
\begin{equation}
f(\omega)\equiv-\frac{\left(  \sum_{q}g_{q}^{2}G_{q}\left(  \omega\right)
e^{iq(R_{1}-R_{2})}\right)  \left(  \sum_{q}g_{q}^{2}G_{q}\left(
\omega\right)  e^{iq(R_{2}-R_{1})}\right)  }{-i(\omega-\omega_{\mathrm{K}%
})+\kappa_{R}/2+i\sum_{q}g_{q}^{2}G_{q}\left(  \omega\right)  }.
\end{equation}
The excitation of the left nanowire propagates to the right nanowire by the
spin waves in the film. When chiral coupling is perfect, $f(\omega)$ vanishes
without the back-action. The microwave output of both left and right nanowires
as inductively detected by coplanar wave guides are denoted $\hat
{p}_{\mathrm{out}}^{(L)}(\omega)$ and $\hat{p}_{\mathrm{out}}^{(R)}(\omega)$
with input-output relations \cite{input_output1,input_output2}
\begin{align}
\hat{p}_{\mathrm{out}}^{(L)}(\omega)  &  =p_{\mathrm{in}}^{(L)}(\omega
)+\sqrt{\kappa_{p,L}}\hat{m}_{L}(\omega),\nonumber\\
\hat{p}_{\mathrm{out}}^{(R)}(\omega)  &  =\sqrt{\kappa_{p,R}}\hat{m}%
_{R}(\omega),
\end{align}
where $\kappa_{p,R}$ is the additional radiative damping induced by the
detector. Therefore, the elements in the microwave scattering matrix
describing reflection $\left(  S_{11}\right)  $ and transmission $\left(
S_{21}\right)  $ amplitudes become
\begin{align}
S_{11}(\omega)  &  \equiv\frac{\hat{p}_{\mathrm{out}}^{(L)}}{\hat
{p}_{\mathrm{in}}^{(L)}}=1-\frac{\kappa_{p,L}}{-i(\omega-\omega_{\mathrm{K}%
})+(\kappa_{L}+\kappa_{p,L})/2+i\sum_{q}g_{q}^{2}G_{q}\left(  \omega\right)
-f(\omega)},\nonumber\\
S_{21}(\omega)  &  \equiv\frac{\hat{p}_{\mathrm{out}}^{(R)}}{\hat
{p}_{\mathrm{in}}^{(L)}}=\left[  1-S_{11}(\omega)\right]  \sqrt{\frac
{\kappa_{p,R}}{\kappa_{p,L}}}\frac{i\sum_{q}g_{q}^{2}G_{q}\left(
\omega\right)  e^{iq(R_{2}-R_{1})}}{-i(\omega-\omega_{\mathrm{K}})+\kappa
_{R}/2+i\sum_{q}g_{q}^{2}G_{q}\left(  \omega\right)  }. \label{S}%
\end{align}
The real parts of $S_{11}$ and $S_{12}$ at different magnetic fields and
microwave frequencies are illustrated in Fig.~\ref{transmission} for
antiparallel magnetizations. The frequency of the Co Kittel mode decreases with increasing
magnetic field until its direction is reversed to the magnetic-field direction
(here $\left\vert H_{\mathrm{app}}\right\vert \lesssim200$~mT). The
interference pattern on the Kittel resonance in Fig.~\ref{transmission}(b)
reflects the transmission phase delay $e^{ik(R_{1}-R_{2})}$ in Eq.~(\ref{S}).
We note that in our model the nanowires do not reflect spin waves, the
features should therefore not be interpreted in terms of standing spin waves.

\subsection{Incoherent chiral pumping}

A temperature gradient between the magnetic nanowire and film also injects
unidirectional magnon currents, i.e., causes a chiral spin Seebeck effect
\cite{Spin_seebeck_exp,Spin_seebeck_theory1,Spin_seebeck_theory2,Spin_caloritronics}%
. Here we consider again two identical transducers, i.e., a magnetic nanowire
at $\mathbf{r}_{2}=R_{2}\hat{\mathbf{y}}$ that detects magnons, which are now
thermally injected by the nanowire at $\mathbf{r}_{1}=R_{1}\hat{\mathbf{y}}$
and $R_{1}<R_{2}$. This is the configuration of the non-local spin Seebeck
effect as detected electrically in many experiments starting with
Ref.~\cite{Ludo}. The magnons in those experiments are believed to be injected
by the interface exchange interaction or generated by a temperature gradient
in the bulk and results are interpreted by spin diffusion models. Here we
consider the regime in which the exchange effect is suppressed, magnon
propagation is ballistic and we disregard the bulk spin Seebeck effect due to
possible temperature gradients. We predict a spin non-local spin Seebeck
effect that is caused exclusively by dipolar fields and carried by magnons
with long wave lengths and lifetimes. We focus on the Kittel magnons in the
wires since the dipolar coupling between the film and higher bands in the
nanowire is very small. The coupling strength $|g_{\mathbf{k}}|$ in
Fig.~\ref{coupling} illustrates that magnons with wavelength around half of
the nanowire width (here $\pi/w=0.045$~nm$^{-1}$) dominate the coupling.
Pumping from other than the those modes can therefore be disregarded even at
elevated temperatures. Furthermore, the spin current in the film is dominated
by spin waves with small momentum and long mean-free paths, so the effects of
magnon-magnon and magnon-phonon interactions that otherwise render magnon
transport phenomena diffuse \cite{Ludo} should be negligibly small. The
narrow-band thermal injection requires an inductive (or optical) detection of
the magnons accumulated in the detector contact, since the inverse spin Hall
effect with heavy metal contacts is very inefficient.

The equation of motions of the Kittel modes in the nanowire and film spin
waves with momentum $q$ in the coupled system read
\begin{align}
\frac{d\hat{m}_{L}}{dt}  &  =-i\omega_{\mathrm{K}}\hat{m}_{L}-\sum_{q}%
ig_{q}^{\ast}e^{iqR_{1}}\hat{\alpha}_{q}-\frac{\kappa}{2}\hat{m}_{L}%
-\sqrt{\kappa}\hat{N}_{L},\nonumber\\
\frac{d\hat{m}_{R}}{dt}  &  =-i\omega_{\mathrm{K}}\hat{m}_{R}-\sum_{q}%
ig_{q}^{\ast}e^{iqR_{2}}\hat{\alpha}_{q}-\frac{\kappa}{2}\hat{m}_{R}%
-\sqrt{\kappa}\hat{N}_{R},\nonumber\\
\frac{d\hat{\alpha}_{q}}{dt}  &  =-i\omega_{q}\hat{\alpha}_{q}-ig_{q}%
e^{-iqR_{1}}\hat{m}_{L}-ig_{q}e^{-iqR_{2}}\hat{m}_{R}-\frac{\kappa_{q}}{2}%
\hat{\alpha}_{q}-\sqrt{\kappa_{q}}\hat{N}_{q},
\end{align}
where $\kappa$ is caused by the same Gilbert damping in both nanowires, and
$\hat{N}_{L}$ and $\hat{N}_{R}$ represent the thermal noise in the left and
right nanowires, with $\langle{\hat{N}_{\eta}^{\dagger}(t)\hat{N}%
_{\eta^{\prime}}(t^{\prime})}\rangle=n_{\eta}\delta(t-t^{\prime})\delta
_{\eta\eta^{\prime}}$. Here, $\eta\in\{L,R\}$ and $n_{\eta}=1/\left\{
\exp\left[  \hbar\tilde{\omega}_{\mathrm{K}}/(k_{B}T_{\eta})\right]
-1\right\}  $ and $T_{R}$ is also the film temperature. Integrating out the
spin-wave modes in the film, we obtain equations for dissipatively coupled
\cite{Yao_Yu,Magnon_radiation} nanowires. In frequency space,
\begin{align}
&  \left(  -i(\omega-\omega_{\mathrm{K}})+\frac{\kappa}{2}+\frac{\Gamma
_{1}+\Gamma_{2}}{2}\right)  \hat{m}_{L}(\omega)+\Gamma_{2}e^{iq_{\ast}%
|R_{2}-R_{1}|}\hat{m}_{R}(\omega)\\
&  =\sum_{q}ig_{q}^{\ast}e^{iqR_{1}}\sqrt{\kappa_{q}}G_{q}(\omega)\hat{N}%
_{q}(\omega)-\sqrt{\kappa}\hat{N}_{L}(\omega),\nonumber\\
&  \left(  -i(\omega-\omega_{\mathrm{K}})+\frac{\kappa}{2}+\frac{\Gamma
_{1}+\Gamma_{2}}{2}\right)  \hat{m}_{R}(\omega)+\Gamma_{1}e^{iq_{\ast}%
|R_{2}-R_{1}|}\hat{m}_{L}(\omega)\nonumber\\
&  =\sum_{q}ig_{q}^{\ast}e^{iqR_{2}}\sqrt{\kappa_{q}}G_{q}(\omega)\hat{N}%
_{q}(\omega)-\sqrt{\kappa}\hat{N}_{R}(\omega), \label{equations_noise}%
\end{align}
where $\Gamma_{1}=|g_{q_{\ast}}|^{2}/v_{q_{\ast}}$ and $\Gamma_{2}%
=|g_{-q_{\ast}}|^{2}/v_{q_{\ast}}$ are assumed constant (for the Kittel mode).
Here, $q_{\ast}$ is the positive root of $\omega_{q_{\ast}}=\tilde{\omega
}_{\mathrm{K}}$.

For perfectly chiral coupling with $\Gamma_{2}=0$ the solutions of
Eqs.~(\ref{equations_noise}) read
\begin{align}
\hat{m}_{L}(\omega)  &  =\frac{\sum_{q}ig_{q}^{\ast}e^{iqR_{1}}\sqrt
{\kappa_{q}}G_{q}(\omega)\hat{N}_{q}(\omega)-\sqrt{\kappa}\hat{N}_{L}(\omega
)}{-i(\omega-\omega_{\mathrm{K}})+\frac{\kappa}{2}+\frac{\Gamma_{1}}{2}%
},\nonumber\\
\hat{m}_{R}(\omega)  &  =\frac{\sum_{q}ig_{q}^{\ast}e^{iqR_{2}}\sqrt
{\kappa_{q}}G_{q}(\omega)\hat{N}_{q}(\omega)-\sqrt{\kappa}\hat{N}_{R}
(\omega)-\Gamma_{1}e^{q_{\ast}(R_{2}-R_{1})}\hat{m}_{L}(\omega)}%
{-i(\omega-\omega_{\mathrm{K}})+\frac{\kappa}{2}+\frac{\Gamma_{1}}{2}}.
\end{align}
With $\hat{m}_{L,R}(t)=\int e^{-i\omega t}\hat{m}_{L,R}(\omega)d\omega/(2\pi
)$, the Kittel modes are occupied according to
\begin{align}
\rho_{L}  &  \equiv\langle\hat{m}_{L}^{\dagger}(t)\hat{m}_{L}(t)\rangle
=n_{L}+\int\frac{d\omega}{2\pi}\frac{\kappa}{(\omega-\omega_{\mathrm{K}}
)^{2}+(\kappa/2+\Gamma_{1}/2)^{2}}(n_{q_{\ast}}-n_{L}),\\
\rho_{R}  &  \equiv\langle\hat{m}_{R}^{\dagger}(t)\hat{m}_{R}(t)\rangle
=n_{R}+\int\frac{d\omega}{2\pi}\frac{\Gamma_{1}^{2}\kappa}{\left[
(\omega-\omega_{\mathrm{K}})^{2}+(\kappa/2+\Gamma_{1}/2)^{2}\right]  ^{2}%
}(n_{L}-n_{q_{\ast}}),
\end{align}
where the damping in the high-quality film has been disregarded $\left(
\kappa_{q}\rightarrow0\right)  $. In the linear regime the non-local thermal
injection of magnons into the right transducer by the left one then reads
\begin{align}
\delta\rho_{R}  &  =\left\{
\begin{array}
[c]{c}%
\mathcal{S}_{\mathrm{CSSE}}(T_{L}-T_{R})\\
0
\end{array}
\text{ when }%
\begin{array}
[c]{c}%
T_{L}>T_{R}\\
T_{L}\leq T_{R}%
\end{array}
\right.  ,\nonumber\\
\mathcal{S}_{_{\mathrm{CSSE}}}  &  =\int\frac{d\omega}{2\pi}\frac{\Gamma
_{1}^{2}\kappa}{\left[  (\omega-\omega_{\mathrm{K}})^{2}+(\kappa/2+\Gamma
_{1}/2)^{2}\right]  ^{2}}\left.  \frac{dn_{L}}{dT}\right\vert _{T=\left(
T_{L}+T_{R}\right)  /2}.
\end{align}
where we defined the chiral (or dipolar) spin Seebeck coefficient
$\mathcal{S}_{_{\mathrm{CSSE}}}.$

The device therefore operates as a heat diode, apparently acting as a
\textquotedblleft Maxwell demon\textquotedblright\ that rectifies the thermal
fluctuations at equilibrium. However, in thermal equilibrium all right and
left moving magnons are eventually connected by reflection of spin waves at
the edges and absorption and re-emission by connected heat baths. The Second
Law of thermodynamics is therefore safe, but it might be interesting to search
for chirality-induced transient effects.

\section{Conclusion and outlook}

Handedness or chirality of wave propagation is a popular research topic in
optics, acoustics, and condensed matter physics. Here we contribute by a
theory for the coherent and incoherent chiral pumping of spin waves into thin
magnetic films through the chiral magnetodipolar radiation generated by the
Oersted field of metallic striplines and dipolar field of magnetic
nanostructures. Spin waves excited coherently in the film under magnetic
resonance of the nanowire are unidirectional, generating a non-equilibrium
magnetization in only half of the film. A temperature gradient between a local
magnet and a film leads to the unidirectional excitation of incoherent
magnons, i.e., a chiral spin Seebeck effect.

\textit{$\mathcal{PT}$} symmetry has been predicted to amplify unidirectional
response \cite{PT_symmetry1,PT_symmetry2,PT_symmetry3}. Even though our system
is dissipative and therefore not PT symmetric, the nonreciprocal coupling of
the two wires still allows directional amplification
\cite{accumulation1,accumulation2}. It would be interesting to introduce
\textit{$\mathcal{PT}$} symmetry into our system via gain in one wire that
compensates the damping in the other, possibly leading to enhanced effects.

Magnons can interact remotely by their chiral dipolar magnetic fields with
other quasiparticles including other magnons, photons, phonons, and conduction
electron spins. Strong chiral coupling between magnons and photons exist,
e.g., in microwave waveguides or cavities that contain chains of small magnets
on special lines \cite{accumulation1,accumulation2}. Large magnon numbers
accumulate at one edge of a chain of magnets when excited by local antennas
\cite{accumulation1,accumulation2}. Spin currents by electrons or phonons may
be generated by the chiral magnetodipolar radiation as well. Chirality is a
functionality that has not yet been employed much in spintronics, but could be
the basis for a new generation of spin-based devices made from conventional materials.

\begin{acknowledgement}
This work is financially supported by the Nederlandse Organisatie voor Wetenschappelijk Onderzoek (NWO) as well as JSPS KAKENHI Grant No. 19H006450. We thank Yaroslav M. Blanter, Haiming Yu, Bi-Mu Yao, Toeno van der Sar, Sanchar Sharma, Yu-Xiang Zhang, Weichao Yu, and Xiang Zhang for helpful discussions.
\end{acknowledgement}

%
%
%

\end{document}